\documentstyle[pra,aps,amssymb,epsf]{revtex}
\begin{document}
\draft
\title{Optically Driven Qubits in Artificial Molecules}
\author{
J. H. Oh\thanks{Electronic address; jungoh@iquips.uos.ac.kr},
D. Ahn\thanks{Also with Department of Electrical Engineering,
University of Seoul, Seoul 130-743, Korea}\thanks{Electronic address;
dahn@uoscc.uos.ac.kr}, 
and 
S. W. Hwang\thanks{Permanent address: Department of Electronics Engineering,
 Korea University, 5-1 Anam, Sungbook-ku, Seoul 136-701, Korea}}
\address{Institute of Quantum Information Processing and Systems\\
University of Seoul, 90. Jeonnong, Tongdaemoon-ku\\
Seoul 130-743, Republic of Korea}
\maketitle
\begin{abstract}
We present novel models of quantum gates based on coupled quantum dots in which
a qubit is regarded as the superposition of ground states in each dot.
Coherent control on the qubit is performed by both a frequency and a polarization of
a monochromatic light pulse illuminated on the quantum dots.
We also show that a simple combination of two single qubit gates functions
as a controlled NOT gate resulting from an electron-electron interaction.
To examine the decoherence of quantum states, we discuss electronic relaxation 
contributed mainly by LA phonon processes.
\end{abstract}

\section{Introduction}
Since useful algorithms such as exhaustive search\cite{Grover} and
factorization\cite{Shor}
have been proposed, there has been considerable interest in quantum gates 
which are the basic units for quantum
computation and information processing.
Quantum gates manipulate quantum states through the unitary transformation with an
externally driven signal and it was shown that two kinds of a quantum gate are enough
for quantum computational procedure; a single-bit and two-bit gates.\cite{DiVincenzo,Barenco}
A single-bit gate controls a single-particle state, called a qubit which is
the superposition of two orthogonal states namely $\mid\!0\rangle$ and $\mid\!1\rangle$;
mathematically a qubit is written as $\mid\psi\rangle=
\cos\theta e^{i\varphi} \mid\!0\rangle+\sin\theta\mid\!1\rangle$.
Thus,
with controllable parameters $\theta$ and $\varphi$,
$\mid\psi\rangle$ spans the entire Hilbert space of 
$\mid\!0\rangle$ and $\mid\!1\rangle$ by a unitary transformation, including a
reversible NOT operation; $\mid\psi\rangle\rightarrow\mid\!{\psi_r}\rangle=
\cos\theta e^{i\varphi}\mid\!1\rangle +\sin\theta\mid\!0\rangle $
on the single-bit gate.
A two-bit gate (controlled NOT gate) functions on two single-qubits called
a target bit $\mid\!\psi\rangle$
and a control bit $\mid\!\chi\rangle=\cos\theta_c e^{i\varphi_c}\mid\!0\rangle_c+
\sin\theta_c \mid\!1\rangle_c$ and evolves the target bit conditionally, {\it i.e.},
depending on the state of the control bit. In detail, if the control bit is 
$\mid\!1\rangle_c$, the gate performs the reversible NOT operation on the target bit.
Otherwise, there is no change.
Hence, for a initial state of $\mid\!\psi\rangle\mid\!\chi\rangle$, we obtain,
by the controlled NOT operation,
\begin{eqnarray}
\mid\!\psi\rangle \mid\!\chi\rangle \longrightarrow 
\cos\theta_c e^{i\varphi_c}
\mid\!\psi\rangle\mid\!0\rangle_c +
\sin\theta_c
\mid\!{\psi_r}\rangle\mid\!1\rangle_c
\label{CNOT}
\end{eqnarray}
where the parameters $\theta_c$ and $\varphi_c$ characterize the state of the control bit.

Various quantum gates have been explored for realizing a quantum computer,
including trapped ions,\cite{Cirac} spins on a nuclear,\cite{Chuang}
and cooper-pairing states in Joshepson box.\cite{Makhlin}
Especially, quantum bits based on the semiconductor quantum dots draw attention relative to
others because advance in nanosemiconductor technology makes it possible to tailor
the quantum dots.  Both spin and spatial parts of wavefunctions for an electron confined
in quantum dot can be exploited as a qubit.
As shown by several authors,\cite{Loss,Burkard}
up-and down-spin states of an electron were found to be a good basis 
for a qubit which  is controlled by a magnetic
field.
It was also shown that the operation of the controlled NOT gate
is easily modeled using the Heisenberg exchange interaction
between two-electron spins.

The spatial part of an electronic wavefunction is also very interesting for a qubit
because its energy can be tailored easily by gate electrodes. 
Moreover, a recent experiment succeeded in controlling the spatial part of wavefunctions
by a light pulse.\cite{Bonadeo}
A typical model along this scheme uses the superposition of the ground and
the first excited states of a single quantum dot as a qubit.
However, since electrons at the excited
state are relaxed rapidly to the ground state by the phonon process,
the coherent time of the case is estimated to be shorter than that of the spin state.
More advanced model is proposed by Openov\cite{Openov} recently where the superposition of each
ground states of two separated quantum dots is viewed as a qubit.\cite{Barenco}
Since two quantum dots are assumed to interact each other only by their excited states,
one expects that an electron at the ground states has longer coherence time
than that of the spin case once it is defined.
It is also shown that by optical illumination, an electron can transfer between dots 
coherently and one can complete the reversible NOT operation with appropriate 
frequency and strength of the light pulse.
However, even though such a qubit is very feasible, models of  single-bit and
two-bit gates are not suggested and detailed discussion for the coherent time is still lacking.

In this work, we suggest novel models of quantum gates based on coupled quantum dots or artificial
molecules. By solving the time-dependent Schr\H{o}dinger equation, we show that
simple artificial molecules associated with light pulses serve as the quantum gates.
In addition, by calculating the relaxation rate by phonon process, 
we also discuss the decoherence of a quantum state during the unitary operation.
For the quantum dots, we consider two-dimensional or disk-like shape with the lateral size
much larger than the extent in the growth direction ($z$-direction) by
patterning isolated metallic gates or etching vertically quantum wells.\cite{Ashoori,VanHove}
Since the size of the quantum dot is comparable to the effective Bohr radius
of a host semiconductor, discrete energy levels are formed in the quantum dot where
the number of electrons and confinement potential are controlled artificially.
Thus, since the number of quantum states in the quantum dot depends on the confinement potential
or the radius of the quantum dot, we assume that one or two levels are bound at each 
isolated quantum dot; the ground and first excited states.
Furthermore, in our case, since the quantum dot is circular symmetric about $z$-axis,
the ground state has an angular momentum $l= 0$($s$ state)
and the first excited state with $l=\pm 1$($p$ states) is degenerate.

\section{Single-bit gate}
In our model, the single-bit gate consists of two larger quantum dots and one smaller dot
which are embedded in a barrier material with a potential energy $V_b$ as shown in Fig. 1.
Depending on its radius, each dot is assumed to have the different number of
atomic states with itself if isolated from others. 
Two larger quantum dots named $A$ and $B$
have both the ground ($l=0$) and the first excited states ($l=\pm 1$) with energies
$\epsilon_s$ and $\epsilon_p$ respectively while a smaller quantum named $C$
has only the ground state $l=0$ with a energy $\epsilon_s^\prime$.
Further, the energies $\epsilon_p$ and $\epsilon_s^\prime$ of the atomic states
at the dot $A$, $B$, and $C$ are assumed to lie close to the barrier potential energy $V_b$.
As a result, wavefunctions of these states have a large extension spatially and the quantum dots
can interact each other via these states to be an artificial molecule.
However, the atomic ground states of the dot $A$ and $B$
are supposed to have  deep energies and be well-isolated from states at the other quantum dots.
This means that an excess electron occupied at the ground states has a large coherence time
as discussed by Openov.\cite{Openov}
For this reason,
we exploit the atomic ground states of the dots $A$ and $B$ as the basis of a qubit,
{\it i.e.} with the ground states for the excess electron in the system of Fig. 1,
 we write a qubit as,
\begin{eqnarray}
   \mid\!\psi\rangle = \cos\theta e^{i\varphi} \mid\!0\rangle+\sin\theta\mid\!1\rangle
   = \cos\theta e^{i\varphi} \mid\!0;A\rangle+\sin\theta\mid\!0;B\rangle
\label{eq:qubit}
\end{eqnarray}
where $\mid\!0\rangle\equiv\mid\!0;A\rangle$, $\mid\!1\rangle\equiv\mid\!0;B\rangle$, 
and $\mid\!l;\alpha\rangle$ is a state located at the quantum dot $\alpha$
with an angular momentum $l$.

Then, single-particle states of the artificial molecule in Fig. 1 are determined
by the Hamiltonian $H_0$,
\begin{eqnarray}
    H_0 &=& \epsilon_s(d_{0A}^\dagger d_{0A}+d_{0B}^\dagger d_{0B})
           +\epsilon_p(d_{\sigma A}^\dagger d_{\sigma A}+d_{\sigma B}^\dagger d_{\sigma B}+
          d_{\pi A}^\dagger d_{\pi A}+ d_{\pi B}^\dagger d_{\pi B}
           +d_{0 C}^\dagger d_{0 C} ) \nonumber \\
        & &+\{Vd_{0C}^\dagger d_{\sigma A}+Vd_{0C}^\dagger d_{\sigma B} +{\rm H. c.}\}
\label{eq:H0}
\end{eqnarray}
where $\epsilon_s^\prime=\epsilon_p$ for simplicity and
$d_{l\alpha}$($d_{l\alpha}^\dagger$) is the electron annihilation(creation) operators
for an electron in states $\mid l;\alpha\rangle$ ($\alpha=A,B,C; l=0,\pm 1$).
Here, operators $d_{\alpha\sigma}$ and $d_{\alpha\pi}$ describe
two orthogonal states localized at the dot $\alpha$ and are defined as
    \begin{eqnarray}
    d_{\sigma\alpha}&=&\frac{-1}{\sqrt{2}}
                      ( e^{ i\phi_\alpha}d_{1\alpha}+e^{-i\phi_\alpha}d_{-1\alpha} )
                      \nonumber \\
    d_{\pi\alpha}   &=&  \frac{-i}{\sqrt{2}}
                      ( e^{-i\phi_\alpha}d_{1\alpha}-e^{ i\phi_\alpha}d_{-1\alpha} )
    \end{eqnarray}
with $\phi_\alpha$ an angle between the $x$-axis and the dot $\alpha$.
These states have $p$-like wavefunctions and, as shown in Fig. 1-(b),
the $\mid\!\sigma;\alpha\rangle$ states have their globes of wavefunctions
directed to the dot C from the dot $\alpha$ whereas
those of the $\mid\!\pi;\alpha\rangle$
 states are normal to the line connecting  the dot C with the dot $\alpha$.
The last term of Eq. (\ref{eq:H0}) represents the interaction between 
excited states of the quantum dots.
Due to a geometircal distance, atomic states of the dots $A$ and $B$
are supposed not to be coupled
to each other, but to be coupled to the ground state of the dot $C$ with
a strength $V$.
It is noted that $\mid\!\pi;\alpha\rangle$ is not coupled to $\mid\!0;C\rangle$
because of a geometrical symmetry.
The Hamiltonian $H_0$ is easily diagonalized and its molecular orbitals $\mid\!k\rangle$
are given as,
    \begin{eqnarray}
\mid 0 \rangle &=& \mid 0;A\rangle,~~ 
\mid 1 \rangle  =  \mid 0;B\rangle,~~\nonumber \\
\mid 2 \rangle &=&\frac{1}{2}(
\sqrt{2}\mid 0;C\rangle -\mid \sigma;B\rangle -\mid \sigma;A\rangle),~~
\nonumber \\
\mid 3 \rangle &=& \frac{1}{\sqrt{2}}( \mid \sigma;B\rangle -\mid \sigma;A\rangle),~~~
\mid 4 \rangle  = \mid \pi;A\rangle,~~
\mid 5 \rangle  =  \mid \pi;B\rangle,~~ \nonumber \\
\mid 6 \rangle &=& \frac{1}{2}(
\sqrt{2}\mid 0;C\rangle +\mid \sigma;B\rangle+\mid \sigma;A\rangle)
\label{eq:newbasis}
    \end{eqnarray}
with their eigenenergies $\epsilon_0\!=\!\epsilon_1\!=\!\epsilon_s$,
$\epsilon_2\!=\!\epsilon_p-\sqrt{2}V$, $\epsilon_3\!=\!\epsilon_4\!=\!\epsilon_5\!=\!\epsilon_p$,
and $\epsilon_6\!=\!\epsilon_p+\sqrt{2}V$. 
As noted previously, both $\mid 0;A\rangle$ and $\mid 0;B\rangle$ are still eigenstates of
the artificial molecule in  Fig. 1 and
the qubit of Eq. (\ref{eq:qubit}) is not evolved without a perturbation.

In order to evolve the qubit, we use optical transitions between molecular states
$\mid\!k\rangle$ of Eq. (\ref{eq:newbasis}).
Under the classical field approximation, an excess electron
interacts with a light by a Hamiltonian
$H_{light}(t)$,
    \begin{eqnarray}
    H_{light}(t)= -\frac{e}{2m^*c}(\vec{A}\cdot\vec{p}+ \vec{p}\cdot\vec{A} )
\label{eq:light0}
    \end{eqnarray}
where $\vec{A}$ is the vector potential of the light,
$\vec{p}$ is a momentum operator of the electron, and  the $\mid\vec{A}\mid^2$ term
is omitted for simplicity.
If the light propagates along the $z$-axis and its electric field is polarized by an angle
$\phi$ from the $x$-axis, the vector potential $\vec{A}$ is given by,
    \begin{eqnarray}
    \vec{A} = \frac{cE_0}{\omega}(\hat{x}\cos\phi+\hat{y}\sin\phi)\cos(\omega z/c-\omega t)
\label{eq:vector}
    \end{eqnarray}
with an electric field strength $E_0$ and  a frequency $\omega$.
Hence, at the $z$-plane on which the quantum dots reside,
the optical transitions of the ground states to or from excited states of the molecule
are governed by,
    \begin{eqnarray}
H_{light}(t) &=& \frac{i(\epsilon_p-\epsilon_s)E_0\xi}{\sqrt{2}\hbar\omega} \cos(\omega t) \Big[
\cos(\phi_B+\phi) d_0^\dagger d_2-\cos(\phi_B-\phi) d_1^\dagger d_2 \nonumber \\
& & +\sqrt{2}\cos(\phi_B+\phi) d_0^\dagger d_3 +\sqrt{2}\cos(\phi_B-\phi) d_1^\dagger d_3
 -2\sin(\phi_B+\phi) d_0^\dagger d_4\nonumber \\
& & -2\sin(\phi_B-\phi) d_1^\dagger d_5
-\cos(\phi_B+\phi) d_0^\dagger d_6 +\cos(\phi_B-\phi) d_1^\dagger d_6 \Big]+{\rm H. c.}
\label{eq:light}
    \end{eqnarray}
where $\xi\equiv\langle 0;\alpha\mid ex\mid 1;\alpha\rangle$ is
the electric dipole moment and the optical transitions between excited states are omitted.

To study the evolution of the wavefunction caused by the illumination of the light,
we write the wavefunction of the excess electron as,
    \begin{eqnarray}
\mid\!\Psi(t)\rangle = \sum_{k=0}^{6} S_k (t) \exp\{-i \epsilon_k t/\hbar\}\mid k\rangle.
\label{eq:gPsi}
    \end{eqnarray}
Here, the expansion coefficient $S_k$ denotes the probability amplitude of the state $\mid\!
k\rangle$ and is determined by the time-dependent Sch\H{o}dinger equation,
    \begin{eqnarray}
i\hbar\frac{\partial \mid\!\Psi(t)\rangle}{\partial t} = H(t) \mid\!\Psi(t)\rangle
\label{eq:tdSch}
    \end{eqnarray}
with $H=H_0+H_{light}(t)$, and
the values of $S_0(0)$ and $S_1(0)$ define an initial qubit with
$S_k(0)=0~(k=2,3,4,5,6)$.

For an effective control of the excess electron, we exploit the optical transition
through the first excited state $\mid\!2\rangle$ 
by tuning the frequency $\omega$ 
much more closer to $(\epsilon_2-\epsilon_0)/\hbar$
than $(\epsilon_n-\epsilon_0)/\hbar$ ($n=3,4,5,6$).
Then, the ground and the first excited states are nearly resonant with the light
while the higher excited states are out of resonance.
In this work, to a good approximation,
we neglect the non-resonant optical transitions
for the range of the frequency,
    \begin{eqnarray}
\mid\hbar\delta\mid \ll \epsilon_3-\epsilon_2=\sqrt{2}V
\label{eq:resonant}
    \end{eqnarray}
with $\delta\equiv(\epsilon_2-\epsilon_0)/\hbar-\omega$. Then,
the Hamiltonian $H$ governing the wavefunction of Eq. (\ref{eq:gPsi})
is replaced with $H_r$,
    \begin{eqnarray}
H_r =\sum_{k=0}^6\epsilon_k d_k^\dagger d_k+
\Big\{ \frac{i\hbar\Omega}{4}e^{i\omega t}\Big[
u d_0^\dagger d_2-v d_1^\dagger d_2\Big]+{\rm H.c.}\Big\}
\label{eq:reduce}
    \end{eqnarray}
with $u=\cos(\phi_B+\phi)$, $v=\cos(\phi_B-\phi)$, and 
 $\hbar\Omega\equiv\sqrt{2}E_0\xi(\epsilon_p-\epsilon_s)/(\epsilon_2-\epsilon_0)$.

The simple form of the Hamiltonian $H_r$ in Eq. (\ref{eq:reduce}) makes it possible
to obtain an analytic solution of the wavefunction Eq. (\ref{eq:gPsi}).
To do this, first we make use of the unitary transformation,\cite{Galitskii}
    \begin{eqnarray}
U(t) =\exp\{ i\omega t/2(d_0^\dagger d_0+d_1^\dagger d_1-d_2^\dagger d_2 ) \}.
\label{eq:evol}
    \end{eqnarray}
Substituting $\mid\!\Psi(t)\rangle = U(t) \mid\!\tilde{\Psi}(t)\rangle$
into the Schr\H{o}dinger equation Eq. (\ref{eq:tdSch}),
we obtain another equation for $\mid\!\tilde{\Psi}(t)\rangle$:
    \begin{eqnarray}
i\hbar\frac{\partial \mid\!\tilde{\Psi}(t)\rangle}{\partial t}
 = \tilde{H}\mid\!\tilde{\Psi}(t)\rangle
\label{eq:tdSch1}
    \end{eqnarray}
with the rotated Hamiltonian $\tilde{H}$
    \begin{eqnarray}
\tilde{H} &=& U^\dagger(t)H_rU(t)-i\hbar U^\dagger(t)\frac{\partial U(t)}{\partial t}
\nonumber \\
&=& (\epsilon_0+\hbar\omega/2)(d_0^\dagger d_0+d_1^\dagger d_1)
+(\epsilon_2-\hbar\omega/2)d_2^\dagger d_2 +\sum_{k=3}^6\epsilon_k d_k^\dagger d_k\nonumber \\
& & + \Big\{ \frac{i\hbar\Omega}{4}\Big[
u d_0^\dagger d_2-v d_1^\dagger d_2\Big]+{\rm H.c.}\Big\}.
    \end{eqnarray}
Since the Hamiltonian $\tilde{H}$ is independent on time,
the solution for $\mid\!\tilde{\Psi}(t)\rangle$ is given by;
    \begin{eqnarray}
\mid\!\tilde{\Psi}(t)\rangle = \exp[-i\tilde{H}t/\hbar]\mid\!\tilde{\Psi}(0)\rangle.
\label{mform}
\end{eqnarray}
Then, as soon as the light pulse is illuminated at the instance $t=0$,
the probability amplitude  $S_k(t)$ in Eq. (\ref{eq:gPsi}) evolves  as,
\begin{eqnarray}
S_k(t) = \sum_{m,n=0}^2 e^{i\epsilon_k t/\hbar}\langle k\!\mid U(t)\mid\! m\rangle
\langle\!m\mid\exp[-i\tilde{H}t/\hbar]\mid\!n\rangle S_n(0), ~~~~k=0,1,2
\label{eq:solution}
\end{eqnarray}
and $S_k(t)=0~~(k=3,4,5,6)$.
Since the evolution operator $\exp[-i\tilde{H}t/\hbar]$ is diagonal for
eigenvectors of the Hamiltonian $\tilde{H}$, it is calculated,
on the basis $\mid\!k\rangle~(k=0,1,2)$, as
    \begin{eqnarray}
\langle\!k^\prime&\!\mid&\!\exp[\frac{( 2\tilde{H} -\epsilon_0-\epsilon_2 )t}{2i\hbar}]
\!\mid\!k\rangle =
\frac{e^{i\delta t/2}}{u^2+v^2}
     \left( 
       \begin{array}{ccc}
       v^2 & u v  & 0 \\
       u v & u^2  & 0 \\
       0   & 0    & 0 \\
       \end{array}
     \right)
    +
\frac{\cos(\Omega_t t/2)}{u^2+v^2}
     \left( 
       \begin{array}{ccc}
       u^2 & -u v  & 0 \\
      -u v & v^2  & 0 \\
       0   & 0    & u^2+v^2 \\
       \end{array}
     \right) \nonumber \\
   &+&
\frac{\sin(\Omega_t t/2)}{2\Omega_t(u^2+v^2)}
     \left( 
       \begin{array}{ccc}
       2i\delta u^2 & -2i\delta u v  & \Omega u(u^2+v^2) \\
      -2i\delta u v & 2i\delta v^2   &-\Omega  v(u^2+v^2)\\
      -\Omega u(u^2+v^2)& \Omega  v(u^2+v^2)   &-2i\delta(u^2+v^2)  \\
       \end{array}
     \right) 
\label{eq:expH}
\end{eqnarray}
with $\Omega_t(\delta) = \sqrt{\delta^2+\Omega^2(u^2+v^2)/4}$. 
Substituting 
Eq. (\ref{eq:expH}) into Eq. (\ref{mform}), we can see that
the off-diagonal components of Eq. (\ref{eq:expH}) describes 
the transfer of the electron between quantum states
because the operator $U(t)$ is diagonal on the basis $\mid\!k\rangle$.
Especially, the transfer of the electron from or to the state $\mid\!2\rangle$
is determined by the last matrix of Eq. (\ref{eq:expH}).
Hence the occupancy of the state $\mid\!2\rangle$ is
oscillating proportional to $\sin(\Omega_t t/2)$ during the illumination of the light.
For this reason, we choose the duration $\tau$ of the light pulse to be
$\sin(\Omega_t \tau/2)=0$ or
    \begin{eqnarray}
    \tau = {2 N\pi}/\Omega_t(\delta)
\label{tau}
    \end{eqnarray}
with an integer $N$.
Then, after a pulse is completed, the excess electron still occupies the ground states
of the dots $A$ and $B$, {\it i.e.},
 for $t\geq\tau$, the wavefunction is given by
    \begin{eqnarray}
\mid\Psi(t)\rangle &=&
 e^{-i \epsilon_0 t}\Big[ S_0(\tau) \mid 0\rangle + S_1(\tau) \mid 1\rangle)\Big]
    \end{eqnarray}
where, from Eq. (\ref{eq:solution}),
    \begin{eqnarray}
     \left( 
            \begin{array}{c}
                S_0(\tau)\\
                S_1(\tau)\\
            \end{array}
     \right) 
&=&
{\cal R}(\phi,\delta)
     \left( 
            \begin{array}{c}
                S_0(0)\\
                S_1(0)\\
            \end{array}
     \right)  \\
{\cal R}(\phi,\delta)
&=&
e^{-i\delta\tau/4}
\left(
     \begin{array}{cc}
-\cos(2\Delta\theta)\cos(\frac{\delta\tau}{4})+i\sin(\frac{\delta\tau}{4}) &
\sin(2\Delta\theta)\cos(\frac{\delta\tau}{4}) \\
\sin(2\Delta\theta)\cos(\frac{\delta\tau}{4}) & 
\cos(2\Delta\theta)\cos(\frac{\delta\tau}{4})+i\sin(\frac{\delta\tau}{4}) \\
            \end{array}
     \right)
\label{rotation}
\end{eqnarray}
for an odd integer $N$ with $\cos(\Delta\theta)\equiv u/\sqrt{u^2+v^2}$ and
$\sin(\Delta\theta)\equiv v/\sqrt{u^2+v^2}$.

Thus, varying the polarization or (and) the frequency of the light pulse
we can control the single qubit.
For instance, the relative phase between $\mid\! 0\rangle$ and $\mid\! 1\rangle$ 
in the qubit can be manipulated by the frequency of the light pulse
with the polarization angle $\phi=\phi_B+\pi/2$, {\it i.e.}, $v=0$.
From Eq. (\ref{rotation}),
we find that the operator ${\cal R}(\phi_B+\pi/2,\delta)$ changes
the phase of the state $\mid\!0\rangle$ by $\pi-\delta\tau/2$
relative to the state $\mid\!1\rangle$,
\begin{eqnarray}
 S_0(0)\mid\!0\rangle
+S_1(0)\mid\!1\rangle
\buildrel {\cal R}\over\longrightarrow 
 e^{i(\pi-\delta\tau/2)} S_0(0) \mid\!0\rangle +S_1(0)\mid\!1\rangle.
\end{eqnarray}
So, by varying the tuning frequency $\delta$ in the range of,
    \begin{eqnarray}
-\frac{\Omega \sin(2\phi_B)}{\sqrt{N^2-1}}\leq \delta \leq
\frac{\Omega \sin(2\phi_B)}{\sqrt{N^2-1}},
\label{range}
    \end{eqnarray}
we get the relative phase changing from a zero to $2\pi$
for $N\geq 3$.
This range of the tuning frequency $\delta$ is also small enough to satisfy
the resonant approximation of Eq. (\ref{eq:resonant}) because of 
usually $\hbar\Omega\ll V$.
It is noted in Eq. (\ref{range}) that $\phi_B\neq 0$ for the control of the relative phase
or, in other words, the artificial molecule should be bent on the $z$-plane like a water molecule.

On the other hand, the light pulse described by
${\cal R}(\phi,0)$ [or ${\cal R}(0,\delta)$]
is very useful to control an amount of the excess electron at the quantum dots 
because its rotation matrix is given by,
\begin{eqnarray}
{\cal R}(\phi,0)=
\left(
     \begin{array}{cc}
-\cos(2\Delta\theta)&
\sin(2\Delta\theta) \\
\sin(2\Delta\theta) & 
\cos(2\Delta\theta)\\
            \end{array}
     \right).
\end{eqnarray}
If this light pulse is illuminated on the qubit of
$\mid\!\psi_0\rangle=\cos\theta_0\mid\!0\rangle+\sin\theta_0\mid\!1\rangle$
with a zero relative phase, the operation ${\cal R}(\phi,0)$ transforms $\theta_0$
to $\theta_0+2\Delta\theta+\pi$.
Using this fact, we can prepare an arbitrary qubit of Eq. (\ref{eq:qubit}) by applying
two successive light pulses of ${\cal R}(\phi_B+\pi/2,\delta){\cal R}(\phi,0)$ 
on the excess electron located initially at the state $\mid\!0\rangle$.
Then, we obtain the final qubit 
$\mid\!\psi\rangle$ with 
$\theta=2\Delta\theta+\pi$ and $\varphi=-\delta\tau/2$.
As a special case of the operation ${\cal R}(\phi,0)$,
the light pulse described by ${\cal R}(0,0)$
serves as the reversible NOT operation, as also found by Openov;\cite{Openov}
\begin{eqnarray}
 \mid\!\psi\rangle = S_0(0)\mid\!0\rangle
+S_1(0)\mid\!1\rangle
\buildrel {\cal R}\over\longrightarrow 
\mid\!{\psi_r}\rangle = S_1(0) \mid\!0\rangle +S_0(0)\mid\!1\rangle.
\end{eqnarray}

\section{Controlled NOT gate}

For the controlled NOT operation, we consider the arrangement of the quantum dots
as shown in Fig. 2 where a target(control) bit is regarded as the superposition
of atomic ground states at two larger quantum dots located at the upper(lower) side.
In detail, the first electron "1" expresses the target bit by occupying
the atomic ground states of the dot $F$ and $G$ as,
   \begin{eqnarray}
\mid\!\psi\rangle = 
\cos\theta e^{i\varphi} \mid\!0\rangle_t+\sin\theta \mid\!1\rangle_t =
\cos\theta e^{i\varphi} \mid\!0;F\rangle_1+\sin\theta \mid\!0;G\rangle_1
\label{tbit}
   \end{eqnarray}
and the occupation of the second electron "2" on the dots $J$ and $K$ defines
the control bit as,
   \begin{eqnarray}
\mid\!\chi\rangle = 
\cos\theta_c e^{i\varphi_c} \mid\!0\rangle_c+\sin\theta_c \mid\!1\rangle_c =
\cos\theta_c e^{i\varphi_c} \mid\!0;J\rangle_2+\sin\theta_c \mid\!0;K\rangle_2.
\label{cbit}
   \end{eqnarray}
As the case of a single-bit gate, we assume that the interaction among the larger quantum
dots is mediated via a single ground state of  a smaller quantum dot $I$.
Furthermore, due to a special location of the smaller quantum dot $I$ in Fig. 2,
atomic excited states($l=\pm 1$) of the three larger quantum dots $F,~G,~J$ are coupled to
the ground state of the dot $I$ while the dot $K$ is supposed to be isolated from
others. 
For this system, the single-particle Hamiltonian is given as 
\begin{eqnarray}
    H^C_0 &=& \epsilon_s(
                      d_{0F}^\dagger d_{0F} +d_{0G}^\dagger d_{0G}
                     +d_{0J}^\dagger d_{0J} +d_{0K}^\dagger d_{0K}) \nonumber \\
        & &+\epsilon_p( d_{\sigma F}^\dagger d_{\sigma F}
                       +d_{\sigma G}^\dagger d_{\sigma G}
                       +d_{\sigma J}^\dagger d_{\sigma J}
                       +d_{\sigma K}^\dagger d_{\sigma K}) \nonumber \\
        & &+\epsilon_p( d_{\pi F}^\dagger d_{\pi F}
                       +d_{\pi G}^\dagger d_{\pi G}
                       +d_{\pi J}^\dagger d_{\pi J}
                       +d_{\pi K}^\dagger d_{\pi K})+\epsilon_p  d_{0I}^\dagger d_{0I}\nonumber \\
        & &+\{Vd_{0I}^\dagger d_{\sigma F}+Vd_{0I}^\dagger d_{\sigma G}
        +V^\prime d_{0I}^\dagger d_{\sigma J}+{\rm H. c.}\}
\label{eq:HC0}
\end{eqnarray}
where we denote the coupling strength between $\mid\!\sigma;J\rangle$ and 
$\mid\!0;I\rangle$ as $V^\prime $ which can be different from the value of $V$ between
$\mid\!0;I\rangle$ and $\mid\!\sigma;F,G\rangle$ depending on the dot-dot distance.

In fact, since the system contains two excess electrons representing
the control and the target bits, respectively, 
there is in general the electron-electron interaction. Then, this interaction
may lead to the hybridization of the ground states at each larger quantum dots with other states,
so that the basis of a quantum bit may be no longer defined as the superposition of the
atomic ground states.
To resolve this problem, we incorporate a metal layer into the substrate, {\it i.e.}
below the quantum dots.
Then, since the excess electrons in the quantum dots are screened by charges in the metal
layer, the electron-electron interaction could be short-ranged.
If the distance of the quantum dots to the metal layer is sufficiently small,
it is reasonable to assume that
the Coulomb interaction is neglected for two electrons located at different quantum dots,
respectively. 
Under this condition, we have still four single-particle ground states
whose wavefunctions are localized at each larger quantum dots and the basis of a qubit
used in the single-bit gate is well-defined.

Including the fact that the electron-electron interaction is short-ranged,
the Hamiltonian of the two excess electrons in the controlled NOT gate
of Fig. 2  can be written as,
\begin{eqnarray}
    H^C = H^{C}_0(1)+H^C_0(2)+\frac{1}{2}\sum_{\alpha,i,j,k,l}V_{ijkl}
              d_{i\alpha}^\dagger(1)d_{j\alpha}^\dagger(2) d_{k\alpha}(2) d_{l\alpha}(1)
\label{eq:H2}
\end{eqnarray}
where $\alpha$ runs over all the quantum dots and $i,j,k,$ and $l$ designate
an angular momentum of a state at the quantum dot $\alpha$.
Here, $V_{ijkl}$ is the matrix element of the Coulomb potential
which has a non-zero value only when two electrons occupy a quantum dot simultaneously.
As one expects, the ground states of the Hamiltonian Eq. (\ref{eq:H2})
are just the direct product of two single-particle ground states located at different
larger quantum dots as,
\begin{eqnarray}
\mid\!0;\alpha\rangle_1\mid\!0;\beta\rangle_2 ~~~~
(\alpha\neq\beta ~~~\alpha,\beta ={\rm~larger~dots}).
\label{eq:ground}
\end{eqnarray}
because there is no the Coulomb interaction between them.
Excited states of two excess electrons in the artificial molecule of Fig. 2 are in general
expressed in the linear combination of all
possible states $\mid\!i;\alpha\rangle_1\mid\!j;\beta\rangle_2$
except for the states in Eq. (\ref{eq:ground}) and can be obtained in a numerical way.
However, for low-lying excited states, an analytic form of a wavefunction can be derived.
Since two electrons occupying the same site of the quantum dots require a large charging energy,
to a good approximation, we expect that 
states such as $\mid\!i;\alpha\rangle_1\mid\!j;\alpha\rangle_2$ 
are not hybridized to the low-lying excited states.
Then, through a simple calculation, we find that
several low-lying excited states of two electrons in the controlled NOT gate have a form of
\begin{eqnarray}
    \mid\!0;\alpha\rangle_1\mid\!\bar{\alpha}\rangle_2~~~{\rm or}~~~
    \mid\!\bar{\alpha}\rangle_1\mid\!0;\alpha\rangle_2~~~~~~~\alpha=F,G,J,K.
\label{eq:lows}
\end{eqnarray}
Here, $\mid\!\bar{\alpha}\rangle$ is the lowest excited state of a single particle in
Eq. (\ref{eq:HC0}) if the quantum dot $\alpha$ is not exist in Fig. 2
and is given as, for each dot $\alpha$,
\begin{eqnarray}
\mid\!\bar{K}\rangle  &=& 
         \frac{1}{\sqrt{2\eta^2+4}}\Big\{ \mid\!\sigma;F\rangle
        +\mid\!\sigma;G\rangle +\eta\mid\!\sigma;J\rangle
        -\sqrt{2+\eta^2}\mid 0;I\rangle\Big\}, \nonumber \\
\mid\!\bar{J}\rangle  &=&
         \frac{1}{2}(\mid \sigma;F\rangle
        +\mid \sigma;G\rangle
        -\sqrt{2}\mid 0;I\rangle),\nonumber \\
\mid\!\bar{G}\rangle &=&
         \frac{1}{\sqrt{2\eta^2+2}}\Big\{\mid \sigma;F\rangle
        +\eta\mid\!\sigma;J\rangle
        -\sqrt{1+\eta^2}\mid 0;I\rangle\Big\}, \nonumber \\
\mid\!\bar{F}\rangle &=&
       \frac{1}{\sqrt{2\eta^2+2}}\Big\{\mid\!\sigma;G\rangle
      +\eta\mid\!\sigma;J\rangle
      -\sqrt{1+\eta^2}\mid\!0;I\rangle\Big\}
\label{eq:lows1}
\end{eqnarray}
where their eigenenergies are
$\epsilon_p-\sqrt{2+\eta^2}V$,
$\epsilon_p-\sqrt{2}V$,
$\epsilon_p-\sqrt{1+\eta^2}V$,
and
$\epsilon_p-\sqrt{1+\eta^2}V$,
respectively, with $\eta\equiv {V^\prime }/{V}$.
In Fig. 3, we show the energy spectrum of the two electrons at the controlled NOT gate
resulted from Eqs. (\ref{eq:ground}) and (\ref{eq:lows}).

To complete the controlled NOT operation, we drive a two-particle state by illuminating
a monochromatic light pulse polarized along the $x$-direction.
Using Eqs. (\ref{eq:light0}) and (\ref{eq:vector})
we derive the interaction of an electron in the system of Fig. 2 with the light as,
    \begin{eqnarray}
    H^C_{light} &=& -\frac{i\sqrt{2}(\epsilon_p-\epsilon_s)\xi E_0}{\hbar\omega}
\cos(\omega t)
       \Big[ \cos(\phi_G) d_{0F}^\dagger d_{\sigma F} +\sin(\phi_G) d_{0F}^\dagger d_{\pi F}
       -\cos(\phi_G) d_{0G}^\dagger d_{\sigma G}   \nonumber \\    
       & &+\sin(\phi_G) d_{0G}^\dagger d_{\pi G} 
          -d_{0J}^\dagger d_{\pi J} 
        -\cos(\phi_K) d_{0K}^\dagger d_{\sigma K}  
            +\sin(\phi_K) d_{0K}^\dagger d_{\pi K}
\Big]+{\rm H.c.}
\label{clight}
    \end{eqnarray}
where $\phi_G$($\phi_K$) is an angle between the $x$-axis and the dot $G$($K$).
It is noted that in the dot $J$ the state $\mid\!\sigma;J\rangle$ is inactive optically
 because the polarization of the light is normal to a globe of its wavefunction.
Then, the total Hamiltonian governing the motion of two particles
in the controlled NOT gate becomes
    \begin{eqnarray}
H^C_{total} = H^C+H^C_{light}(1)+H^C_{light}(2)
\label{eq:Hctotal}
    \end{eqnarray}
from Eq. (\ref{eq:H2}).

In order to demonstrate the controlled NOT operation of the Hamiltonian $H^C_{total}$,
we expand the two-particle wavefunction
in terms of states of Eqs. (\ref{eq:ground}) and (\ref{eq:lows}) as,
    \begin{eqnarray}
\mid\!\Psi_2(t)\rangle &=& 
 \exp\{-i H^C t/\hbar\}
\sum_{\alpha}\Big[ \sum_{\beta\neq\alpha} S_{\alpha\beta}(t)
\mid\!0;\alpha\rangle_1\mid\!0;\beta\rangle_2 \nonumber \\
& & +Y_\alpha(t)
\mid\!\bar{\alpha}\rangle_1\mid\!0;\alpha\rangle_2 
+  Z_\alpha(t)
\mid\!0;\alpha\rangle_1\mid\!\bar{\alpha}\rangle_2\Big]
\label{eq:wave2}
    \end{eqnarray}
and then, solve the time-dependent Sch\H{o}dinger equation of the Hamiltonian
$H^C_{total}$ for an initial two-qubit,
$\mid\!\Psi_2(0)\rangle = \mid\!\psi\rangle \mid\!\chi\rangle $ defined in Eqs. (\ref{tbit})
and (\ref{cbit}).
In this case, the initial values of $S_{\alpha\beta}(t)$ are given by,
    \begin{eqnarray}
S_{FJ}(0) &=& \cos\theta e^{i\varphi}\cos\theta_c e^{i\varphi_c},~~~ 
S_{GJ}(0)  =  \sin\theta\cos\theta_c e^{i\varphi_c},\nonumber \\
S_{FK}(0) &=& \cos\theta e^{i\varphi}\sin\theta_c,~~~~~~~~ 
S_{GK}(0)  =  \sin\theta\sin\theta_c
\label{eq:initial}
    \end{eqnarray}
and all the others are zero.
Considering the initial condition that the electron "2" is located at
the ground states of the dots $J$ and $K$ for $t\leq 0$, we can simplify
the problem further.
That is, the electron "2" can not be evolved to the state $\mid\!\bar{\alpha}\rangle$
because both the states $\mid\!0;J\rangle$ and $\mid\!0;K\rangle$ are not coupled
to any excited states of Eq. (\ref{eq:lows}) according to Eq. (\ref{clight}).
So, $Z_\alpha(t) = 0$ and Eq. (\ref{eq:wave2}) is reduced to
    \begin{eqnarray}
\mid\!\Psi_2(t)\rangle &=& 
     \exp\{-2i\epsilon_s t/\hbar\}
    \Big[ S_{FK}(t)\mid\!0;F\rangle_1
         +S_{GK}(t)\mid\!0;G\rangle_1
         +Y_{ K}(t) e^{-i\omega_K t} \mid\!\bar{K}\rangle_1\Big]\mid\!0;K\rangle_2+ \nonumber \\
    & & 
 \exp\{-2i\epsilon_st/\hbar\}
\Big[ S_{FJ}(t)\mid\!0;F\rangle_1
     +S_{GJ}(t)\mid\!0;G\rangle_1
     +Y_{ J}(t) e^{-i\omega_J t}\mid\!\bar{J}\rangle_1\Big]\mid\!0;J\rangle_2
\label{eq:wave22}
    \end{eqnarray}
where $\hbar\omega_J = \epsilon_p-\epsilon_s-\sqrt{2}V$ and 
      $\hbar\omega_K = \epsilon_p-\epsilon_s-\sqrt{2+\eta^2}V$.
Thus, we can see that, depending on the state of the electron "2", the optical
transition of the electron "1" occurs via $\mid\!\bar{K}\rangle_1$ or $\mid\!\bar{J}\rangle_1$.

Now, for the controlled NOT gate, we tune the frequency of the light  equal to $\omega_K$
corresponding to the energy difference between the ground state of Eq. (\ref{eq:ground})
and the first excited state $\mid\!\bar{K}\!\rangle_1\mid\!{0;K}\!\rangle_2$.
Then, the state $\mid\!\bar{J\!}\rangle_1\!\mid\!0;J\!\rangle_2$
is out of resonance  and the optical transition to it is suppressed 
because its energy is larger
than that of  the first excited state by $(\sqrt{2+\eta^2}-\sqrt{2})V$.
Under the resonant approximation, we neglect the optical transition
between the ground state and the state $\mid\!\bar{J\!}\rangle_1\!\mid\!0;J\!\rangle_2$.
As a result, only in the case of the electron "2" at the state $\mid\!0;K\rangle_2$,
the electron "1" evolves and its motion is determined by the Hamiltonian
$H^C_r$ projected from $H^C_{total}$,
    \begin{eqnarray}
H^C_r &=& \epsilon_s(d_{0F}^\dagger d_{0F}+d_{0G}^\dagger d_{0G}) +(\epsilon_s+\hbar\omega_K)
d_{\bar{K}}^\dagger d_{\bar{K}} \nonumber \\
&+&
\frac{\hbar\Omega_C}{\sqrt{8}}\Big\{ie^{i\omega_K t}\Big[
d_{0F}^\dagger d_{\bar{K}}-d_{0G}^\dagger d_{\bar{K}}\Big]+{\rm H. c.} \Big\}
    \end{eqnarray}
with
    \begin{eqnarray}
\Omega_C =\frac{\sqrt{2}(\epsilon_p-\epsilon_s)\xi E_0\cos(\phi_G)}
{\hbar^2\omega_K\sqrt{2+\eta^2}}.
    \end{eqnarray}
Since the above Hamiltonian $H^C_r$ is similar to Eq. (\ref{eq:reduce}),
through the same procedure done in the case of the single-bit gate, we obtain
the expansion coefficients of Eq. (\ref{eq:wave2}) as a function of a time,
    \begin{eqnarray}
     \left( 
            \begin{array}{c}
                S_{FK}(t)\\
                S_{GK}(t)\\
                Y_{K}(t)\\
            \end{array}
     \right) 
=
\left( \begin{array}{ccc}
       \cos^2(\Omega_C t/4) & \sin^2(\Omega_C t/4) & \frac{1}{\sqrt{2}}\sin(\Omega_C t/2) \\
       \sin^2(\Omega_C t/4) & \cos^2(\Omega_C t/4) &-\frac{1}{\sqrt{2}}\sin(\Omega_C t/2) \\
      -\frac{1}{\sqrt{2}}\sin(\Omega_C t/2) & \frac{1}{\sqrt{2}}\sin(\Omega_C t/2) 
 &\cos(\Omega_C t/2) \\
            \end{array}
     \right)  
     \left( 
            \begin{array}{c}
                S_{FK}(0)\\
                S_{GK}(0)\\
                Y_{K}(0)\\
            \end{array}
     \right).
\end{eqnarray}
Hence, if we choose the duration of the light pulse as,
    \begin{eqnarray}
    \tau_C = 2 N \pi /\Omega_C,
    \end{eqnarray}
we obtain the wavefunction of Eq. (\ref{eq:wave2}) for $t\geq\tau_C$ as
    \begin{eqnarray}
\mid\!\Psi_2(t)\rangle &=& 
    \Big[ S_{GK}(0)\mid\!0;F\rangle_1
         +S_{FK}(0)\mid\!0;G\rangle_1 \Big]\mid\!0;K\rangle_2 \nonumber \\
   &+&
\Big[ S_{FJ}(0)\mid\!0;F\rangle_1
     +S_{GJ}(0)\mid\!0;G\rangle_1 \Big]\mid\!0;J\rangle_2 \nonumber \\
 &=&
    \Big[ S_{GK}(0)\mid\!0\rangle_t
         +S_{FK}(0)\mid\!1\rangle_t \Big]\mid\!1\rangle_c 
    + 
\Big[ S_{FJ}(0)\mid\!0\rangle_t
     +S_{GJ}(0)\mid\!1\rangle_t \Big]\mid\!0\rangle_c
    \end{eqnarray}
in which we use the fact that $S_{FJ}(t)$, $S_{GJ}(t)$, and $Y_{J}(t)$ are independent of time.
Substituting the initial condition of Eq. (\ref{eq:initial}), we can see that
the above equation represents exactly the controlled  NOT operation of
Eq. (\ref{CNOT}).

\section{Decoherence of quantum states}

So far, we assume that the line-width broadening of a state in a quantum dot is zero,
{\it i.e.} an electronic life time at the state is infinite because
the line width is proportional to the scattering rate.
In reality, the level of the state is broadened due to various mechanisms 
such as impurities, structural imperfection, and phonons.
Especially, the broadening from phonons is important for the practical 
quantum gates\cite{Ahn}
because it is inherent to the solid-state device while others can be controlled
by improved technology.
The scattering of an electron from both longitudinal-acoustic(LA)
and longitudinal-optic(LO) phonons has been studied extensively for low-dimensional
systems such as quantum wells, wire, and dots.\cite{Bockelmann}
In quantum wells, the dominant scattering of the electron results from the LO phonons
via the Fr\H{o}hlich interaction. In a quantum dot, however, this process is forbidden
due to the discrete nature of the levels, unless the level separation equals to
the LO phonon energy $\hbar\omega_{LO}$.\cite{Inoshita}

Now, we examine the electron-phonon scattering mainly contributed by LA phonons.
The scattering rate is calculated in first-order perturbation theory using
the Fermi golden rule,
    \begin{eqnarray}
\Gamma = \frac{2\pi}{\hbar}\sum_{f,\vec{q}} M(q)^2
\mid\!\langle\psi_f\mid e^{i\vec{q}\cdot\vec{r}}
\!\mid\psi_i\rangle\mid^2 
\delta(E_f-E_i\pm \hbar\omega_q)
[ n_B+\frac{1}{2}\pm\frac{1}{2} ]
\label{srate}
    \end{eqnarray}
where the upper(lower) signs account for emission(absorption) of phonons by the transition
of an electron from an initial state $\mid\!\psi_i\rangle$ to
a final state $\mid\!\psi_f\rangle$. The sum extends over all possible
final quantum states $\mid\!\psi_f\rangle$ and phonon wave vector $\vec{q}$.
$n_B$ stands for the Bose distribution function $n_B = [e^{\hbar\omega_q/kT}-1]^{-1}$
with $\omega_q=c_s q$ and a longitudinal velocity of sound $c_s$.
For a given deformation potential $\Xi$, the coupling strength of the electron to LA phonons
is given by
    \begin{eqnarray}
M(q)^2 = \frac{\Xi^2}{2\rho c_s\Omega_v}\hbar q
    \end{eqnarray}
with a mass density $\rho$ and a system volume $\Omega_v$.
To calculate the quantity $\langle\psi_f\mid e^{i\vec{q}\cdot\vec{r}}\!\mid\psi_i\rangle$,
we use wavefunctions as,
    \begin{eqnarray}
\langle\vec{r}\mid\!~~0;\alpha\rangle &=& 
\Big(\frac{1}{\pi^3\lambda_p^4\lambda_z^2}\Big)^{1/4}
e^{-(x^2+y^2)/2\lambda_p^2-z^2/2\lambda_z^2} \nonumber \\
\langle\vec{r}\mid\!\pm 1;\alpha\rangle &=& 
\frac{x\pm i y}{\lambda_p}
\langle\vec{r}\mid\!0;\alpha\rangle
    \end{eqnarray}
which are eigenstates of the quantum dot with a parabolic confinement potential proportional
to $(x^2/\lambda_p^2+y^2/\lambda_p^2+ z^2\lambda_p^2/\lambda_z^4)$. Here, we choose
$\lambda_p\gg\lambda_z$ to model a disk-like quantum dot.

Since the electron occupies the ground states or the first excited state
in the quantum gate during the quantum operation, it is important to examine the
scattering from those states.
First,
in the case of the electron initially at the ground state $\mid 0;\alpha\rangle$,
the scattering occurs to the excited state $\mid n\rangle$
of Eq. (\ref{eq:newbasis}) by absorbing phonons and its rate is given as
    \begin{eqnarray}
\Gamma &=& \sum_{n} P_n \Gamma_0(q_n) \nonumber \\
\Gamma_0(q) &=& \frac{\Omega_v}{4\pi\hbar^2c_s}q^2 M(q)^2 n_B(q) \int_{-1}^1
du \exp\{-q^2\lambda_p^2(1-u^2)-q^2\lambda_z^2 u^2\} (1-u^2)
\label{rate0}
    \end{eqnarray}
where $q_n = (\epsilon_n-\epsilon_0)/\hbar c_s$ and $P_n$ is the probability of
finding $\mid\!\sigma;\alpha\rangle$ or $\mid\!\pi;\alpha\rangle$ in
the excited state $\mid\!n\rangle$.
For a GaAs quantum dot with parameters $\lambda_p=100\AA$ and $\lambda_z=20\AA$,
we show $\Gamma_0(q)$ in Fig. 4 with a solid line as a function
of an energy difference $E=\epsilon_n-\epsilon_0$.
We find that the scattering rate strongly depends on the energy difference $E$.
Especially, for a high energy $E\geq 2meV$,
the scattering is estimated to be very rare.
This means that, if the energy difference between the first excited and the ground states
in the quantum gates is sufficiently large and not close to $\hbar\omega_{LO}$,
the electron at the ground state is rarely scattered to the excited states and has
a long coherent time in the quantum gates.
 
Through a similar argument, we can see that the electron initially at the first excited
states is not relaxed to the ground state by emitting phonons, but frequently
scattered to adjacent excited states by absorbing phonons because their energy difference
is relatively small.
For example, the electron at the first excited state in the single-bit gate
is scattered to the second one with the rate of,
    \begin{eqnarray}
\Gamma_2 = \frac{\Omega_v q^6\lambda_p^4}{8\pi\hbar}M(q)^2 n_B(q)
\int_{-1}^1 du \exp\{-q^2\lambda_p^2(1-u^2)-q^2\lambda_z^2 u^2\} (1-u^2)^2
    \end{eqnarray}
where $q = (\epsilon_3-\epsilon_2)/\hbar c_s$.
Plotting the result as a function of the energy difference $E=\epsilon_3-\epsilon_2$
as shown in Fig. 4,
we can see that the electron at the first excited state is scattered frequently
over a more wide range of $E$ than
that at the ground states.
If the coupling strength $V$ is $1meV$ or $E=\sqrt{2}meV$,
the electron of the first excited state
is scattered with the rate of $10^{11}/sec$.
Therefore, for the single-bit gate to work well, the duration $\tau$
of a light pulse should be smaller than the inverse of the scattering rate 
$\Gamma_2$,
{\it i.e.},
    \begin{eqnarray}
\frac{2N\pi}{\Omega\cos(\phi_B)/\sqrt{2}} \ll \frac{1}{\Gamma}~~~{\rm or}~~~
\Omega \gg N\Gamma.
    \end{eqnarray}
For a GaAs dot, $\hbar\Omega$ should have a larger value than $0.4meV$ at 300 K 
for $\Gamma=10^{11}/sec$
or the field strength of the light $E_0\gg 400V/cm$ for a 10nm disk-like dot
from Eq. (\ref{tau}).

In summary, we present novel models of quantum gates based on coupled quantum dots
or artificial molecules.
By varying the size and the location of each quantum dot in the artificial molecule,
we assume that well-localized ground states are present while  
excited states form molecular orbits with a particular geometrical symmetry.
First, for the single-bit gate, we locate a smaller quantum dot between two
larger quantum dots as shown in Fig. 1. Since two larger quantum dots in Fig. 1
are separated enough,
their ground states are well-localized and we define a qubit as
the superposition of them.
To drive the qubit, we exploit the optical transition between
the ground states and the first excited state of the artificial molecule.
Since the wavefunction of the first excited state extends over all three quantum dots
arranged with  a "V" shape,
we show that, by solving the time-dependent Sch\H{o}dinger equation, 
the light pulse rotates the qubit coherently depending on its frequency and 
polarization to demonstrate the single-bit operation.
Secondly, for the controlled NOT operation, we examine the artificial molecule shown
in Fig. 2 where two larger quantum dots containing a control bit are added to
the single-bit gate with a target bit.
Under the illumination of the light pulse, we show that
the electron representing the target bit evolves conditionally, {\it i.e.}
depending on the state of the control bit because of the asymmetrical location of 
two added quantum dots.
Furthermore, when the electron-electron interaction is short-ranged, we find that
the light pulse with the resonant frequency between the ground and the first excited states
severs as the controlled NOT operation.
Finally, to examine the decoherence of quantum states, we discuss electronic relaxation 
contributed mainly by LA phonon processes.
By calculating the scattering rate using the Fermi-golden rule, we estimate
the duration and the field strength of the light pulse.
\acknowledgments{
This work was supported by the Korean Ministry of Science and Technology
through the Creative Research Initiatives Program under Contract No.
98-CR-01-01-A-20.}


\newpage
\begin{center}
{\large Figure Captions}
\end{center}
{\bf Fig}. 1.
An artificial molecule consisted of three disk-like quantum dots
on the substrate is shown for the single-bit gate in (a).
The ground state of each larger quantum dot is assumed to be well-localized and is used 
as the basis of a qubit.
However, excited states such as $\mid\!\sigma;A,B\rangle$ [solid line in (b)]
extend their wavefunctions over the dot $C$ and form molecular states.

\hspace{0.3cm}
\newline
\hspace{0.3cm}
{\bf Fig}. 2.
We show the arrangement of the quantum dots for the controlled NOT gate.
Here, a target(control) bit is regarded as the superposition of the grounds states
at two larger quantum dots located at the upper(lower) side and their interaction is mediated
by the  quantum dot $I$.

\hspace{0.3cm}
\newline
\hspace{0.3cm}
{\bf Fig}. 3.
 We show a schematic structure of two-particle energy levels.
The conjugate state $\mid\!\beta\rangle_1\mid\!\alpha\rangle_2$ to
the state $\mid\!\alpha\rangle_1\mid\!\beta\rangle_2$ is also possible, however,
it is not shown here for simplicity.

\hspace{0.3cm}
\newline
\hspace{0.3cm}
{\bf Fig}. 4.
Log plot of the phonon scattering rate is shown as a function of an energy
difference $E$ between two relevant states.
$\Gamma_0$ ($\Gamma_2$) with a solid (dotted) line describes the scattering rate 
when an electron is initially at
the ground (first excited) state of the single-bit gate by absorbing phonons.
Used parameters are $\Xi=6.8eV$, $\rho=5.36g/cm^3$, and $c_s=5150m/sec$ for GaAs.

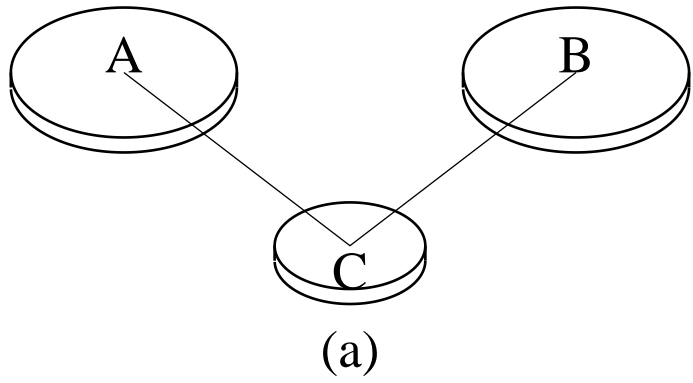
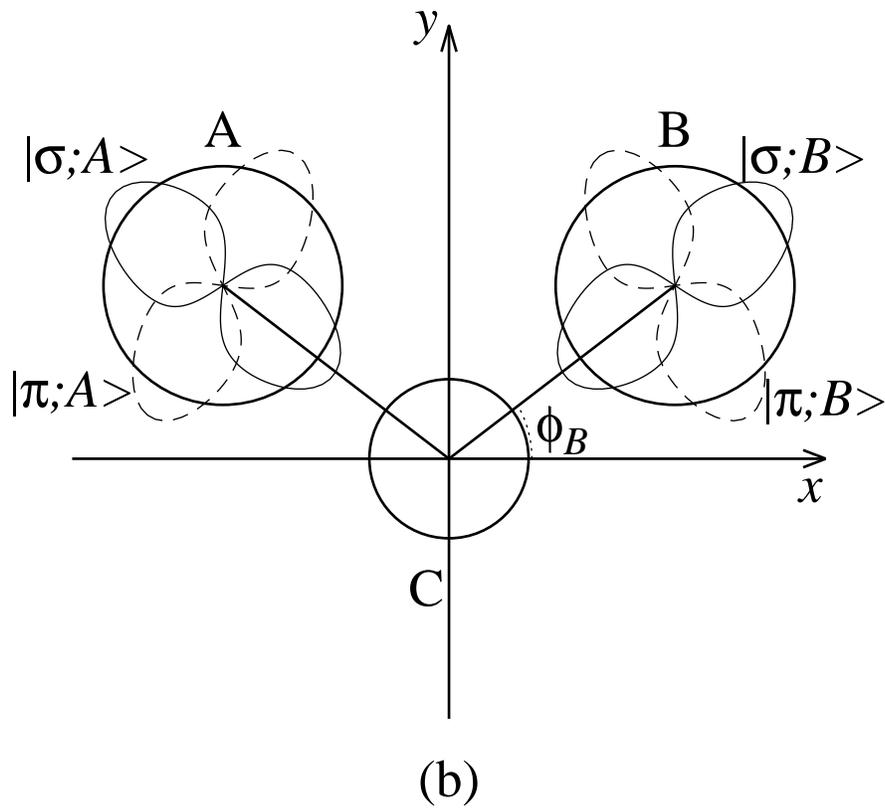
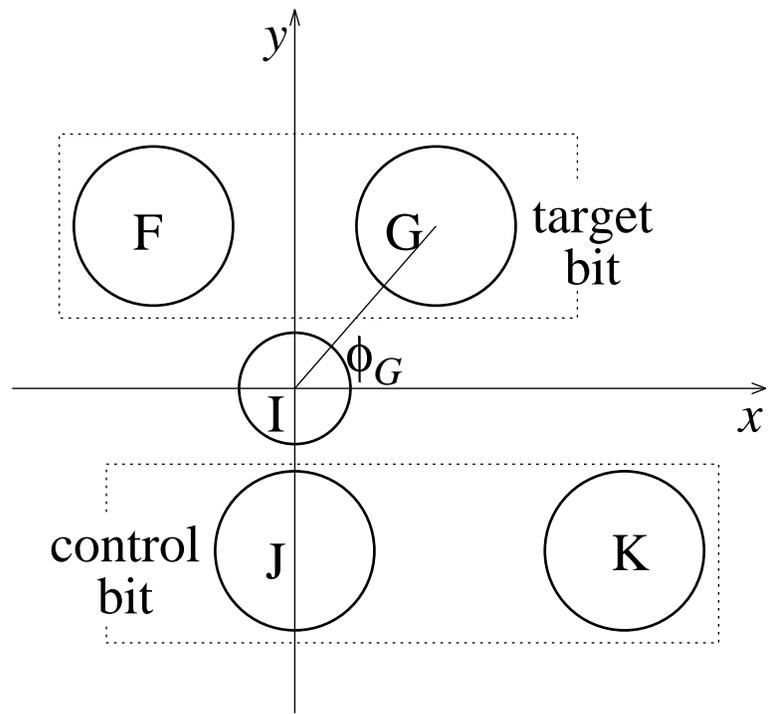
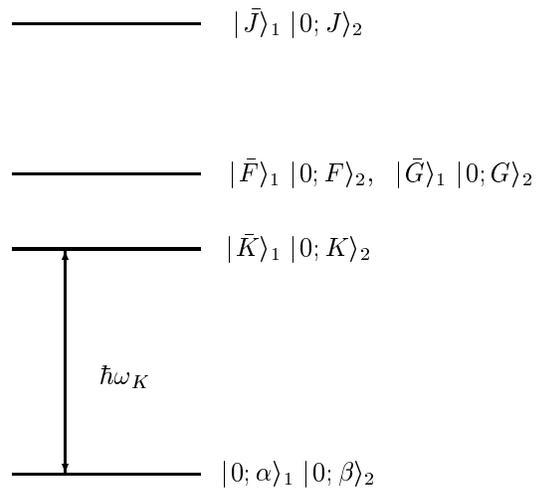
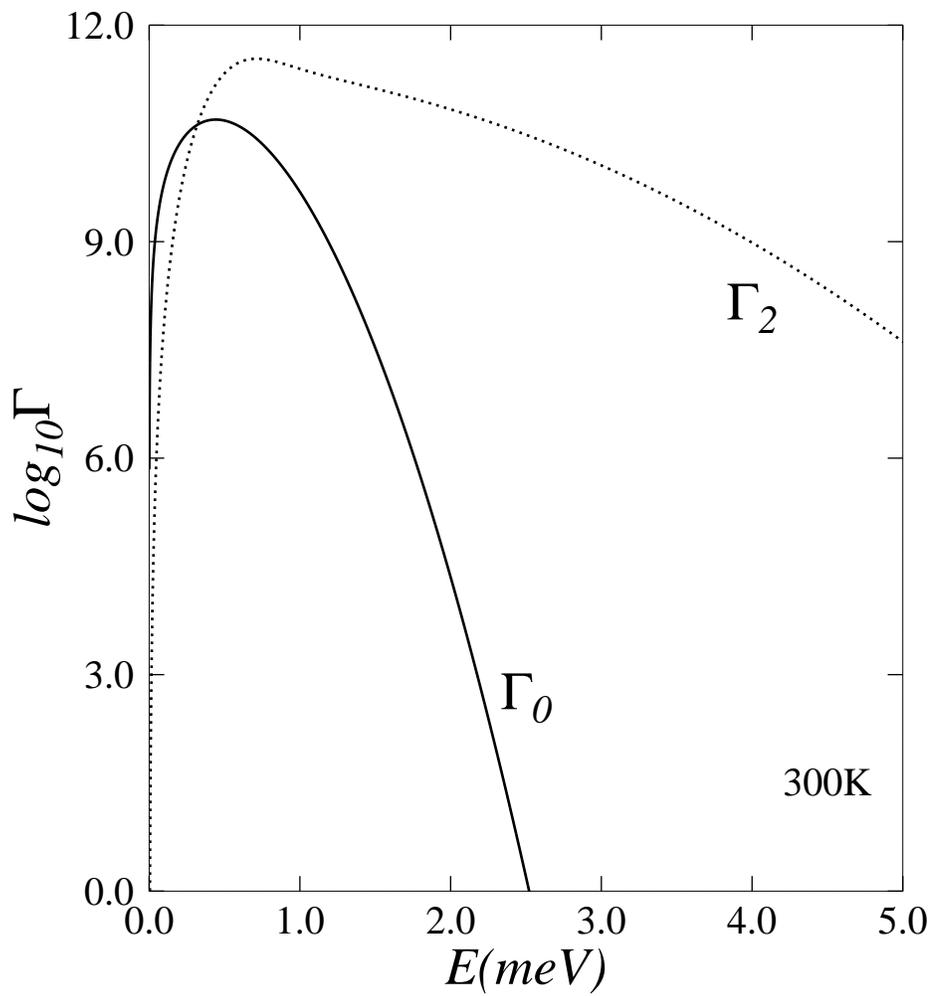
\end{document}